\documentclass[aps,amsmath,amssymb,nofootinbib,superscriptaddress,showpacs,floatfix,prb,twocolumn]{revtex4-1}

\usepackage{latexsym}
\usepackage{graphicx}
\usepackage{times,psfrag,subfigure}
\usepackage{amsmath}
\usepackage{dsfont}
\usepackage{dcolumn}
\usepackage{bm,bbm}       
\usepackage{color}
\usepackage{latexsym,amsmath,amssymb,bm,euscript}
\usepackage{dsfont}
\usepackage{sidecap}

\newcommand{\clw}{\color{white}}

\newcommand{\beq}{\begin{equation}}
\newcommand{\eeq}{\end{equation}}
\newcommand{\beqarray}{\begin{eqnarray}}
\newcommand{\eeqarray}{\end{eqnarray}}

\begin{document}

\allowdisplaybreaks

\title{Helical Majorana surface states of strongly disordered topological superconductors 
\\ with time-reversal symmetry}
\date{\today}

\author{Raquel Queiroz}
\email{r.queiroz@fkf.mpg.de}
\affiliation{Max-Planck-Institut f\"ur Festk\"orperforschung,
  Heisenbergstrasse 1, D-70569 Stuttgart, Germany}

\author{Andreas P. Schnyder}
\email{a.schnyder@fkf.mpg.de}
\affiliation{Max-Planck-Institut f\"ur Festk\"orperforschung,
  Heisenbergstrasse 1, D-70569 Stuttgart, Germany}

\begin{abstract}
Noncentrosymmetric superconductors with strong spin-orbit coupling and the B phase of ${}^3$He
are possible realizations of topological superconductors with time-reversal symmetry.
The nontrivial topology of these time-reversal invariant superconductors
manifests itself at the material surface in the form of helical Majorana modes.
 In this paper, using extensive numerical simulations, we investigate  
the stability and properties of these Majorana states under strong surface disorder, which influences
both bulk and surface states.
To characterize the effects of strong disorder,
we compute the level spacing statistics and the local density of states 
of both two- and three-dimensional topological superconductors.
The Majorana surface states, which are located in the outermost layers
of the superconductor, are protected against weak  disorder, due to their topological characteristic.
Sufficiently strong disorder, on the other hand,
partially localizes the surface layers, with a more pronounced effect on states with energies close to the gap  than 
on those with energies close to zero. 
In particular, we observe that for all disorder strengths and configurations there always exist two extended states at zero-energy that can carry thermal current.
At the crossover from weak to strong disorder 
 the surface state wave functions and the local density of states show signs of critical delocalization.
We find that at this crossover the edge density of states of two-dimensional topological superconductors
exhibits a zero-energy divergence, reminiscent of the Dyson singularity of quasi-one-dimensional dirty superconductors.
\end{abstract}

\date{\today}


% PACS-No:
% 03.65.vf: Topological phases (quantum mechanics)
% 74.50.+r:  tunneling phenomena (superconductivity)
% 74.20.Rp: Pairing symmetries (superconductivity)
% 74.25.F-: Superconductors: transport properties
% 73.20.-r: Surface states, 
% 85.75.-d: spin polarized transport devices (magnetic devices)
% 73.20.Fz: Anderson localization: surface and interface states

\maketitle

\section{Introduction}

Topological superconductors  are attracting a growing interest, due to fundamental considerations as well as potential use for applications in quantum information   
and device fabrication.\cite{Schnyder09,Kitaev09,Ryu10,nayakRMP08,BeenakkerReview,stanescu2013,volovikLectNotes13,Schnyder08,HasanKaneReview,qiZhangRMP11}  
As a consequence of the bulk-boundary correspondence,
topological superconductors host chiral or helical Majorana states at the material surface.\cite{Schnyder08,HasanKaneReview,qiZhangRMP11,volovikBook03,QiHughesZhangPRL09,royArxiv08}
Noncentrosymmetric superconductors (NCSs)\cite{BauerSigristBook,yipReview14,satoPRB06,Iniotakis07,Schnyder10b,Brydon10,Brydon11,Yada11,Dahlhaus12,Matsuura12,Sato11,vorontsovPRL08,Lu10,Queiroz14,Hofmann13,brydonArXiv2014,brydonNJP2013,SchnyderPRL13,yamakage12,Bauer10,Karki10,Yuan06,Nishiyama07,Bordet1991359,bauerPRL04,Kimura05,Sugitani06,Frigeri04} 
with strong spin-orbit coupling have been proposed as candidate materials for time-reversal invariant topological superconductivity,
and the B~phase of   superfluid ${}^3$He\cite{volovikBook03,nagaiJPSJ08,murakawaNagaiPRL09} is believed to be an experimental realization of a topological superfluid with time-reversal symmetry.
The Majorana states of these systems are protected by time-reversal and particle-hole symmetry and 
exhibit a helical spin texture.
That is, the spin orientation of the surface states is coupled to their momentum. 

While due to  charge neutrality the helical Majorana mode does not couple to electromagnetic fields, it
can be detected by applying an effective gravitational field, i.e., a temperature gradient or rotations of the superconductor.\cite{luttingerPRA64,nomuraNaotoPRL12,nakaiNomuraPRB14}
At zero energy one helical Majorana surface mode carries a quantized thermal conductance, 
which is given by $\kappa / T = \pi k_B^2 / (6 h)$.\cite{grinsteinPRB94,senthilFisherPRB00}
Since there is no relevant or marginal time-reversal symmetric perturbation that can be added to the surface Dirac Hamiltonian,\cite{Foster14,nakaiNomuraPRB14,Evers08}
the helical Majorana state is robust against the influence of  weak disorder and interactions. 
Only variations of the Fermi velocity are allowed as a symmetry preserving deformation of the effective Dirac equation describing the surface state, and hence 
an energy gap cannot be opened in the surface spectrum.
However, this argument is based on an effective description of the surface state, which breaks down for perturbations with a strength $\gamma$ larger than the bulk
superconducting gap $\Delta$. In order to study the effects of disorder with $\gamma \geq \Delta$, the disorder induced coupling between the surface and bulk states needs to be taken into account. 
While the effects of disorder on topological superconductor surface states
have been investigated extensively in terms of low-energy effective Dirac theories,\cite{Foster14,Evers08,xieFoster14,grinsteinPRB94,senthilFisherPRB00}
the study of disordered Majorana states at the surface of bulk lattice models has remained an open problem.

In this paper, we examine the effects of strong disorder on the helical Majorana surface modes of two- and three-dimensional topological superconductors in the symmetry
class~DIII using bulk  lattice Hamiltonians.
This is of  relevance for experiments, since surfaces of unconventional superconductors are often intrinsically disordered, or can be disordered on purpose by depositing impurity atoms
using, for example, sputtering techniques.
To investigate the effects of strong surface impurities on the surface and bulk quasiparticle wave functions, we employ large-scale numerical simulations of two- and three-dimensional
Bogoliubov-de Gennes (BdG) lattice Hamiltonians and compute the local density of states
and the level spacing statistics of the wave functions.  
As a prototypical example of a time-reversal invariant topological superconductor, we consider
a noncentrosymmetric superconductor with ($s+p$)-wave pairing symmetry.
We find that  weak nonmagnetic impurities with $\gamma \ll \Delta$ do not affect the helical Majorana state in any way [Fig.~\ref{band3d}(b)], in agreement
with the fact that there does not exist any marginal or relevant perturbation of the surface Dirac Hamiltonian. For moderate nonmagnetic impurities
with $\gamma \sim \Delta$, the disorder induces a strong coupling between the bulk and surface states. 
 Interestingly, the disorder affects the states with energies close to the gap more strongly than those with energies close to zero.
Indeed, we find that for all disorder strengths
extended zero-energy states exist,
which indicates that a diffusive thermal metal phase cannot be realized at
the surface of a class DIII topological superconductor.~\cite{senthilFisherPRB00}
With increasing disorder strength $\gamma$ the
wave functions in the surface layer become less and less delocalized  and show for $\gamma \sim 5t$ (i.e., for a disorder strength of the order of the band width) a probability density that is reminiscent of critical delocalization.\cite{Chou14}
Very strong disorder with $\gamma \gg t$, on the other hand, induces 
 localization in the surface layer and leads to the reappearance
of linearly dispersive Majorana bands in the second and third inward layers.
That is, for $\gamma \gg t$  the effects of disorder on the superconductor and its surface states effectively become weaker [Fig.~\ref{dosdist3D}].
Finally, we also consider magnetic surface disorder, which
removes the time-reversal symmetry protection of the surface states.
We find that whether the Majorana surface state is gapped out, sensitively depends on the direction 
in which the impurity spins are polarized [Fig.~\ref{band_mag}].

This paper is organized as follows. In Sec.~\ref{sec:zwei}, we introduce the BdG Hamiltonian describing a noncentrosymmetric superconductor with 
nontrivial topology, discuss its symmetry properties, and derive the surface Dirac Hamiltonian. Using exact diagonalization we investigate in Sec.~\ref{sec:drei3D} and Sec.~\ref{sec:2D} the influence of 
nonmagnetic impurities on the surface states of  three-dimensional 
and two-dimensional topological superconductors, respectively.
To this end, we compute the local surface density of states, the momentum-resolved spectral function, and
the level spacing statistics. 
This is followed by a brief discussion of the effects of magnetic surface disorder in Sec.~\ref{sec:mag}.
We conclude in Sec.~\ref{sec:summary} with a summary and an outlook on future work.

\section{Surface states of time-reversal invariant topological superconductors}
\label{sec:zwei}

As a prototypical example of a time-reversal invariant topological superconductor in  class DIII we study two- and three-dimensional 
noncentrosymmetric superconductors.\cite{BauerSigristBook}
Fully gapped noncentrosymmetric superconductors with dominant triplet pairing have been shown to be topologically nontrivial 
both in two and three dimensions.\cite{Sato09,Schnyder10b,Tanaka12}

\subsection{Model definition}
\label{sec:NCS}

At a phenomenological level noncentrosymmetric superconductors can be described by a $4 \times 4$ BdG Hamiltonian 
 $\mathcal{H}=\frac{1}{2}\sum_{\bf k}\Phi^\dag_{\bf k}H_{\bf k}\Phi^{\phantom{\dag}}_{\bf k}$, with 
 \begin{subequations}\label{modelDEF}
\begin{eqnarray}  \label{Ham} 
H_{\bf{k} }
=
\begin{pmatrix}
h_{\bf k}&  
\Delta_{\bf k}  \cr
\Delta^{\dag}_{\bf k}  & -h_{\bf -k}^{\mathrm{T}} \cr   
\end{pmatrix} ,
\end{eqnarray}
and the Nambu spinor $\Phi_{\bf k}=(c^{\phantom{\dag}}_{\bf k\uparrow},c^{\phantom{\dag}}_{\bf k\downarrow},c^\dag_{\bf -k\uparrow},c^\dag_{\bf -k\downarrow})^{\mathrm{T}}$, 
where $c_{{\bf k} s }$ represents the electron annihilation operator with momentum ${\bf k}$ and spin $s$.
The normal part of the Hamiltonian $h_{\bf k}={\varepsilon}_{\bf k} \sigma_0 +\lambda{\bf l_k\cdot \bm{\sigma}}$ describes electrons on a square or cubic lattice
with kinetic term  $\varepsilon_{\bf k}=\sum_{i=1}^d t\cos(k_i) -\mu$. Here, $t$ denotes the nearest-neighbor hopping amplitude, $\mu$ is the chemical potential,
$\bm{\sigma}$ is the vector of Pauli matrices, 
$\lambda{\bf l_k\cdot \bm{\sigma}}$ represents a Rashba-type spin-orbit coupling with strength $\lambda$, and
 $d\in\{2,3\}$ is the spatial dimension. 
Momenta are measured in units of the inverse lattice spacing $a^{-1}$. For convenience we set $a=1$ throughout this paper.
The antisymmetric spin-orbit coupling vector ${\bf l}_{\bf k}$ is restricted by the symmetries of the noncentrosymmetric crystal.
In the following we consider a three-dimensional NCS with cubic point group $O$ as well as a two-dimensional NCS with
tetragonal point group $C_{4v}$.
Within a tight-binding expansion, the lowest order term for the cubic point group $O$ is written as
\begin{align} \label{O_SOC}
{\bf l_k}=\sin k_x {\bf \hat x}+\sin k_y {\bf \hat y}+\sin k_z {\bf \hat z} .
\end{align}
Examples of $O$ point-group NCSs include Mo$_3$Al$_2$C\cite{Bauer10,Karki10}, 
Li$_2$Pd$_x$Pt$_{3-x}$B,\cite{Yuan06,Nishiyama07} 
and La$_3$Rh$_4$Sn$_{13}$.\cite{Bordet1991359}
For the $C_{4v}$ point group, which is relevant for thin films of CePt$_3$Si,\cite{bauerPRL04} 
we have
\begin{align} \label{C4vSOC}
{\bf l_k}=\sin k_y {\bf \hat x}-\sin k_x {\bf \hat y}.
\end{align}
In passing we note that two-dimensional topological superconductors with SOC given by Eq.~\eqref{C4vSOC} can also be engineered in
heterostructures,\cite{sasakiAndoArxiv14,FangKanatzidisPRB14} for example, by depositing a material with strong SOC on a conventional $s$-wave superconductor. 

In the absence of inversion symmetry the superconducting gap function $\Delta_{\bf k}$ contains both spin-singlet and spin-triplet pairing components
\begin{align}
\Delta_{\bf k}=(\Delta_\mathrm{s}\sigma_0+\Delta_\mathrm{t}{\bf d_k\cdot\sigma})(i\sigma_2),
\end{align}
\end{subequations}
where $\Delta_\mathrm{s}$ and $\Delta_\mathrm{t}$ denote the spin-singlet and spin-triplet pairing amplitudes, respectively. 
The triplet pairing vector ${\bf d}_{\bf k}$ is assumed to be aligned with the spin-orbit pseudovector ${\bf l}_{\bf k}$, since this choice
maximizes the superconducting transition temperature.\cite{Frigeri04} Moreover, we take the pairing amplitudes $\Delta_\mathrm{s}$ and $\Delta_{\mathrm{t}}$ to
be real and positive.\cite{gaugeTrafo}
For our numerical calculations, we set the model parameters to
$t=4.0$,
$\lambda=-2.0$,
$\Delta_\mathrm{s}=0.5$,
and
$\Delta_\mathrm{t}=2.0$. For the three-dimensional NCS with SOC~\eqref{O_SOC} we 
set the chemical potential to $\mu = 8.0$, whereas for the two-dimensional NCS with SOC~\eqref{C4vSOC}
we set $\mu=4.0$. 
{ With this parameter choice the two- and three-dimensional boundary states decay into the bulk
with a decay length of the order of three lattice spacings.}
We have checked that our results do not depend on the particular choice of the model parameters as long as the triplet pairing component is dominant, i.e., 
$\Delta_{\mathrm{t}} >  \Delta_{\mathrm{s}}$.

Hamiltonian~\eqref{modelDEF} is invariant under all the symmetries of symmetry class DIII. That is, $H_{\bf k}$ satisfies time-reversal symmetry 
$T^{-1} H_{- {\bf k} } T = + H_{\bf k}$,
with $T^2 = - \mathbbm{1}$,
and particle-hole symmetry 
$C^{-1} H_{- {\bf k}} C = - H_{\bf k}$,
with $C^2 = +\mathbbm{1}$.
The time-reversal and particle-hole symmetry operators are given 
by $T =  \sigma_0 \otimes i \sigma_2 \mathcal{K}$ and $C= \sigma_1 \otimes \sigma_0 \mathcal{K}$, respectively, where $\mathcal{K}$ denotes the complex conjugation operator.
For dominant triplet pairing $\Delta_\mathrm{t} > \Delta_\mathrm{s}$, Hamiltonian~\eqref{modelDEF} is topologically nontrivial both in 
two and three dimensions. In three dimensions, the topological characteristics are described by a three-dimensional winding number ($\mathds{Z}$-type invariant), 
while in two-dimensions the topology is characterized by a $\mathds{Z}_2$ index.\cite{Schnyder08}

% %%%%%%%%%%%%%%%%%%%%%%%%%%%%
\begin{figure*}[t]
\flushleft
\hspace*{-0.1cm}\includegraphics[clip,angle=0,width=2.05\columnwidth]{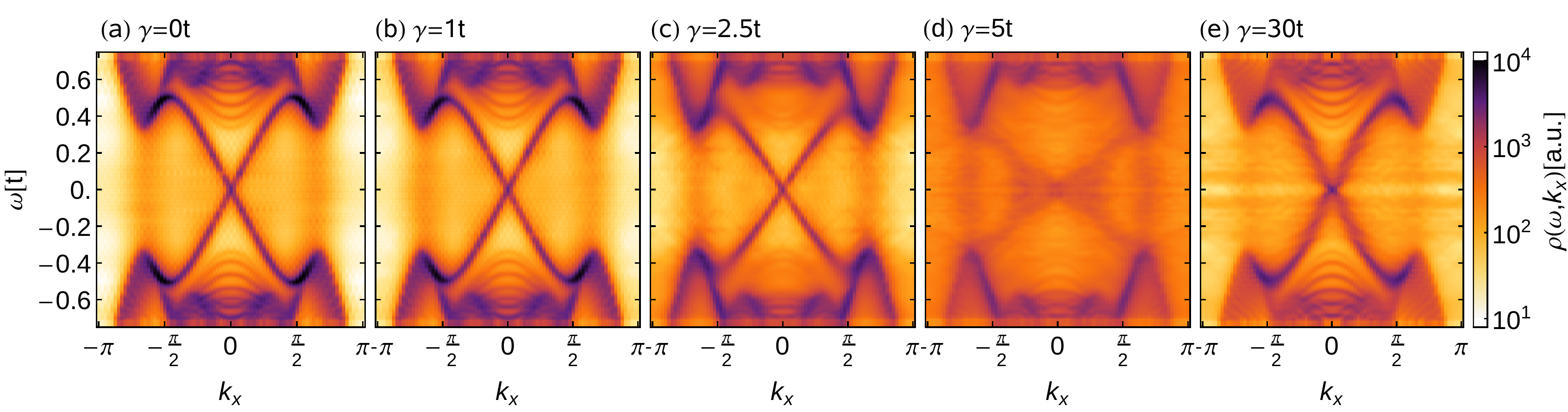}
\caption{\label{band3d}  
(Color online) 
Spectral function $A  ( \omega,  {\bf k}_{\parallel} )$, Eq.~\eqref{spectFun}, on a log scale as a 
function of energy $\omega$ and surface momentum $k_x$ with $k_y=0$ for
the first three outermost layers at the (001) surface of a three-dimensional topological NCS.
The strength $\gamma$ of the Gaussian distributed surface disorder increases from $\gamma = 0t$ in (a) to $\gamma = 30t$ in (e). 
}  
\end{figure*}
%%%%%%%%%%%%%%%%%%%%%%%%%%%%%

\subsection{Effective low energy Hamiltonian}  
\label{lowEnergyHams}

Before studying the effects of surface disorder, let us first derive a low-energy effective Dirac Hamiltonian describing the helical Majorana surface states
both for the three-dimensional and two-dimensional topological NCS.

\subsubsection{Surface states of three-dimensional topological NCS}

In the long-wave-length limit, we can disregard the momentum dependence of the normal part of the Hamiltonian and replace $h_{\bf k}$ by a position-dependent mass term
$m(z)$, with $m(z) \to  + m_0$ for $z \to +\infty$ and $m(z) \to  - m_0$ for $z \to -\infty$,
which describes a domain wall centered at $z=0$. The domain-wall bound states can be written as $e^{i (k_x x + k_y y) }  \psi^{\mathrm{2D}}_{1(2)} ( z)$, with
\begin{eqnarray}
\psi^{\mathrm{2D}}_{1 (2) } ( z) =
\exp
\left[ + \int_0^z d z' \frac{�m ( z' ) }{ \Delta_\mathrm{t} } \right]  \Psi^{\mathrm{2D}}_{1 (2) } , 
\end{eqnarray}
$\Psi^{\mathrm{2D}}_1 =
\frac{1}{\sqrt{2}} (
i, 0 , 0 , 1
)^{\mathrm{T}}
$,
and
$\Psi^{\mathrm{2D}}_2  
=
\frac{1}{\sqrt{2}} ( 0, i , 1 , 0 )^{\mathrm{T}}
$,
where we have neglected the influence of a finite spin-singlet pairing amplitude. 
The low-energy effective Hamiltonian $\widetilde{H}_{\bf k}$ for the surface state is obtained by projecting the Hamiltonian onto
the subspace $\Psi^{\mathrm{2D}} = \left\{ \Psi^{\mathrm{2D}}_1, \Psi^{\mathrm{2D}}_2 \right\}$ spanned by the two bound-state wavefunctions 
$\Psi^{\mathrm{2D}}_1$ and $\Psi^{\mathrm{2D}}_2$, i.e., 
\begin{eqnarray} \label{3DsurfaceHam}
\widetilde{H}^{\mathrm{2D}}_{\bf k} 
=
 \left\langle \Psi^{\mathrm{2D}} \right|  H_{\bf k} \left | \Psi^{\mathrm{2D}} \right\rangle
=
\Delta_{\mathrm{t}}
( k_y \sigma_1 - k_x \sigma_2 ) . \label{2dhamedge}
\end{eqnarray}
Time-reversal and particle-hole symmetry operators in the subspace formed by the surface modes are given by
$\widetilde{T}_{\mathrm{2D}} = - i \sigma_2 \mathcal{K}$ and $\widetilde{C}_{\mathrm{2D}} = - i \sigma_1 \mathcal{K}$, respectively.  
{ Hence, Eq.~\eqref{2dhamedge} describes} a single valley DIII Dirac Hamiltonian.
{ The two symmetries of class DIII} severely restrict the possible perturbations that can be added to the surface Hamiltonian~\eqref{3DsurfaceHam}.\cite{Schnyder08}
We find that the mass term $m \sigma_3$ is prohibited by time-reversal symmetry, whereas the
chemical potential  $\mu \sigma_0$ is forbidden by particle-hole symmetry. Moreover the $U(1)$ gauge potentials
$A_x \sigma_2$ and $A_y \sigma_1$ 
are disallowed since they break both time-reversal and particle-hole symmetry.
The only symmetry-allowed perturbations are variations of the spin-triplet amplitude $\Delta_\mathrm{t}$ which are odd in momentum~${\bf k}$. 
These momentum-dependent perturbations can be neglected in the long-wave-length approximation.

However, the effective description \eqref{3DsurfaceHam} does not capture perturbations with a magnitude larger than 
the gap energy $\Delta_{\mathrm{t}}$. Hence, the helical Majorana states could potentially become unstable in the presence of 
surface disorder with a strength $\gamma > \Delta_{\mathrm{t}}$. We will study this question in detail in Sec.~\ref{sec:drei} using exact diagonalization 
of the bulk Hamiltonian~\eqref{modelDEF}.

\subsubsection{Edge states of two-dimensional topological NCS}
\label{edgeHam2D}

Repeating similar steps as above, we find that the low-energy Hamiltonian describing the edge states of a two-dimensional topological superconductor
in class DIII is given by
\begin{eqnarray} \label{2DsurfaceHam}
\widetilde{H}^{\mathrm{1D}}_{\bf k} 
=
\left\langle \Psi^{\mathrm{1D}} \right| H_{\bf k} \left| \Psi^{\mathrm{1D}} \right\rangle
=
- \Delta_{\mathrm{t}} k_x \sigma_3,
\end{eqnarray}
where $\Psi^{\mathrm{1D}}$ is the space formed by the edge modes, which is spanned by $\Psi^{\mathrm{1D}}_1 = \frac{1}{\sqrt{2}} ( 0, i, 0, 1)^{\mathrm{T}}$ and 
$\Psi^{\mathrm{1D}}_2 = \frac{1}{\sqrt{2}} ( - i, 0, 1, 0)^{\mathrm{T}}$.
Time-reversal and particle-hole symmetry in the edge-mode subspace read
$\widetilde{T}_{\mathrm{1D}} = - i \sigma_2 \mathcal{K}$ and $\widetilde{C}_{\mathrm{1D}} = - i \sigma_3 \mathcal{K}$, respectively. 
As before, we find that there is no symmetry-allowed mass term that can be added to the edge Hamiltonian~\eqref{2DsurfaceHam},
indicating that the helical Majorana edge modes are robust against disorder with strength $\gamma < \Delta_{\mathrm{t}}$.
To demonstrate that the two-dimensional topological NCS, as opposed to the three-dimensional one, has a $\mathbbm{Z}_2$-type topological characteristic,
let us consider a doubled version of the edge Hamiltonian~\eqref{2DsurfaceHam}, i.e., 
$\Delta_t k_x \sigma_3 \otimes \sigma_0$.
In contrast to Hamiltonian~\eqref{2DsurfaceHam}
the doubled Hamiltonian $\Delta_t k_x \sigma_3 \otimes \sigma_0$, can be fully gapped out by the symmetry preserving mass term $m \, \sigma_2 \otimes \sigma_2$.
That is, two-dimensional topological NCS are characterized by an odd number of Kramer's pairs of Majorana edge states.

\subsection{Surface disorder}
\label{sec:SurfDisorder}

 %%%%%%%%%%%%%%%%%%%%%%%%%%%%
\begin{figure}[t!]
%\hspace*{-0.0cm}
\vspace{-0.2cm}
\includegraphics[clip,angle=0,width=1.0\columnwidth]{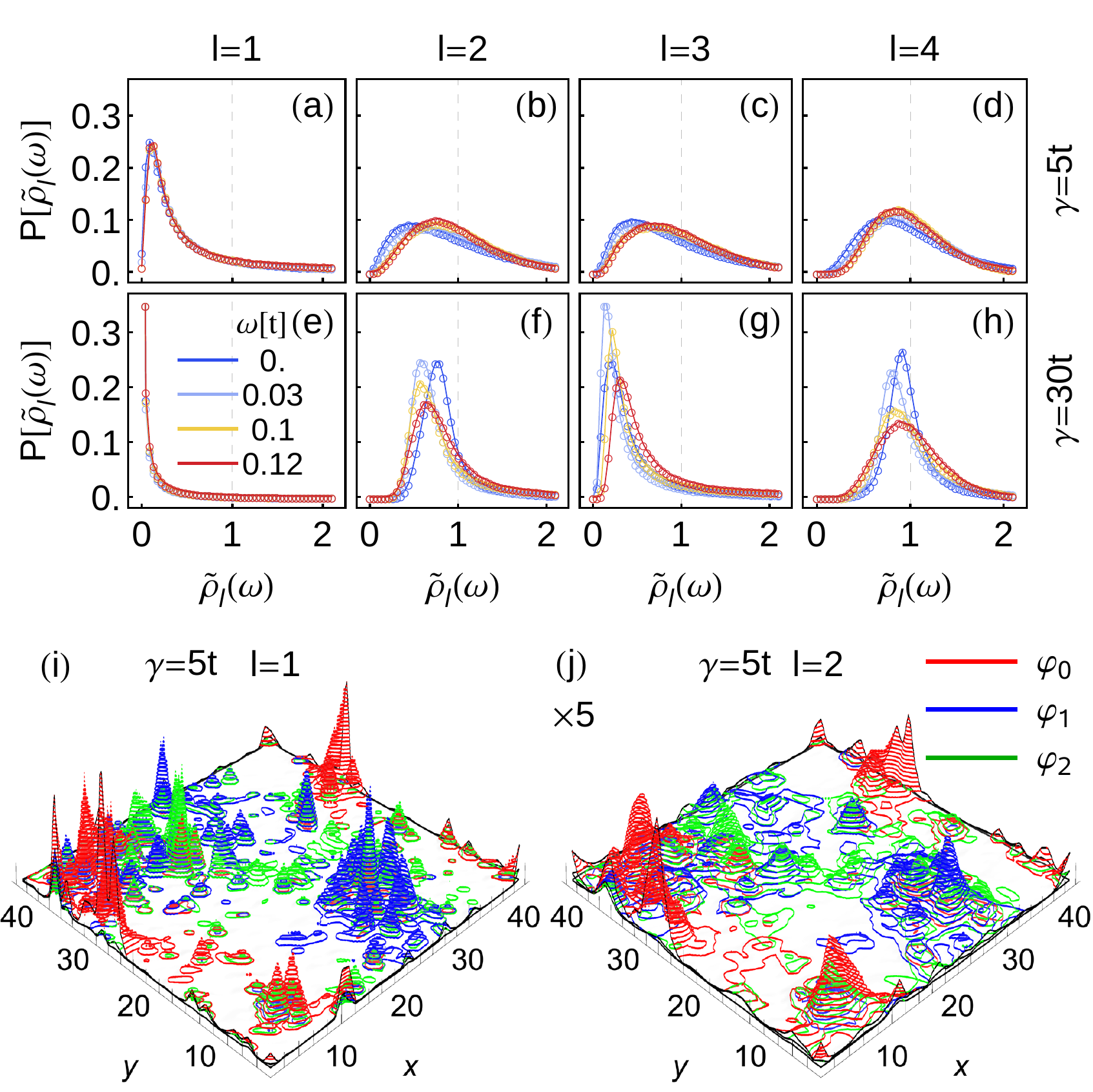}
\caption{\label{dosdist3D}  
(Color online) (a)-(h) Layer and energy-resolved probability distribution $P[ \tilde{\rho}_l ( \omega)]$ of the normalized local density of states $\tilde{\rho}_l (\omega) $
for an ensemble of one hundred disordered three-dimensional topological NCSs, with disorder strengths $\gamma = 5 t$  [panels (a)-(d)] and $\gamma = 30 t$ [panels (e)-(h)]. 
{ Each column shows the probability distribution $P[ \tilde{\rho}_l (\omega)]$ for a different layer, with $l=1$ the surface layer [panels (a) and (e)] and $l=4$ the 4th inward layer [panels (d) and (h)].
The energy dependence is indicated by the color scale, with blue representing $\omega =0$ and red $\omega=\Delta/3$.
Panels (i) and (j) show the wave function probability density $| \bm{\varphi}_m ( l, {\bf r}) |^2$ in the first ($l=1$) and second inward layer ($l=2$), respectively, for the three lowest positive energy wave functions $\bm{\varphi}_0$, $\bm{\varphi}_1$, and $\bm{\varphi}_2$ with
successive eigenvalues $E_0 = 0 < E_1 < E_2$ (red, green, blue), at the disorder strength $\gamma = 5 t$.
We find that at this disorder strength the wave functions show signs of critical delocalization.
The amplitude in panel (j) has been multiplied by a factor of $5$ for clarity. }
$\bm{\varphi}_0$ (red) is the helical Majorana state 
at exactly zero energy.
}
\end{figure}
%%%%%%%%%%%%%%%%%%%%%%%%%%%

To study the stability of the helical Majorana states under strong surface disorder,
we consider uncorrelated random on-site potentials given by
\begin{eqnarray}\label{surfaceDis}
\mathcal{H}^{\beta}_{\mathrm{imp}}
=
\sum_{{\bf k}, {\bf q}} 
\Phi^{\dag}_{\bf k}
V^{\beta}_{\bf q} 
\Phi^{\phantom{\dag}}_{{\bf k}+{\bf q}} ,
\end{eqnarray}
where $V^{\beta}_{\bf q}= (1/ \sqrt{ \mathcal{N} } ) \sum_{n} v ( {\bf r}_n ) S^{\beta} e^{- i {\bf q} \cdot {\bf r_n} }$ denotes the Fourier
transform of the onsite scatterers $v( {\bf r}_n )S^{\beta}$ at the surface sites ${\bf r}_n$ with strength $v ( {\bf r}_n)$.
We investigate the effects of both nonmagnetic impurities ($\beta = 0$)
and magnetic scatterers ($\beta = \{x,y,z \} $)
described by
 $v ( {\bf r}_n ) S^{\beta = 0} = v ( {\bf r}_n )  \sigma_3 \otimes \sigma_0$
 and  $v ( {\bf r}_n ) S^{\beta = \{ x,y,z  \} } = v ( {\bf r}_n ) (\sigma_3\otimes\sigma_1,\sigma_0\otimes\sigma_2,\sigma_3\otimes\sigma_3)$, respectively. 
 The disorder distribution is assumed to be Gaussian like, i.e., for each lattice site on the surface
 the local potential $v ( {\bf r}_n )$ is drawn from a box distribution with $p [�v ({\bf r}_n) ] = 1 / \gamma $ for
 $v ({\bf r}_n) \in \left[ - \gamma /2 , +  \gamma /2 \right]$.
 As discussed above, nonmagnetic disorder with $\gamma < \Delta_{\mathrm{t}}$ does not couple
 to the surface states.  Impurity spins, on the other hand, lift the time-reversal protection of the helical Majorana modes and
 can therefore strongly modify the surface states even for $\gamma < \Delta_{\mathrm{t}}$. 
 Within the low-energy theory of Sec.~\ref{lowEnergyHams}, we find that impurity spins give rise to the following additional 
 term in the low-energy Hamiltonian~\eqref{3DsurfaceHam} describing the helical Majorana modes of the three-dimensional topological NCS
 \begin{eqnarray} \label{eqV3D}
 \left\langle \Psi^{\mathrm{2D}} \right|  V^{\beta}_{\bf q} \left | \Psi^{\mathrm{2D}} \right\rangle 
  =
 \left\{
 \begin{array}{c l}
 v_{\bf q} \sigma_3  &  \quad \textrm{if $\beta =z$} \cr
 0  & \quad  \textrm{otherwise}
 \end{array}
 \right. .
 \end{eqnarray}
That is, only the out-of-plane spin component of magnetic impurities couples to the surface states. 
Similarly, the edge states of the two-dimensional topological NCSs only couple to the $x$ spin component 
of  impurity spins, since
\begin{eqnarray} \label{eqV2D}
 \left\langle \Psi^{\mathrm{1D}} \right|  V^{\beta}_{\bf q} \left | \Psi^{\mathrm{1D}} \right\rangle 
  =
   \left\{
 \begin{array}{c l}
- v_{\bf q} \sigma_1 & \quad \textrm{if  $\beta = x$} \cr
 0 & \quad \textrm{otherwise} 
 \end{array} 
 \right. .
\end{eqnarray}

\section{Numerical results} \label{sec:drei}

  %%%%%%%%%%%%%%%%%%%%%%%%%%%%
\begin{figure}[t]
\vspace{0.25cm}
%\hspace*{-0.0cm}
\includegraphics[clip,angle=0,width=1.02\columnwidth]{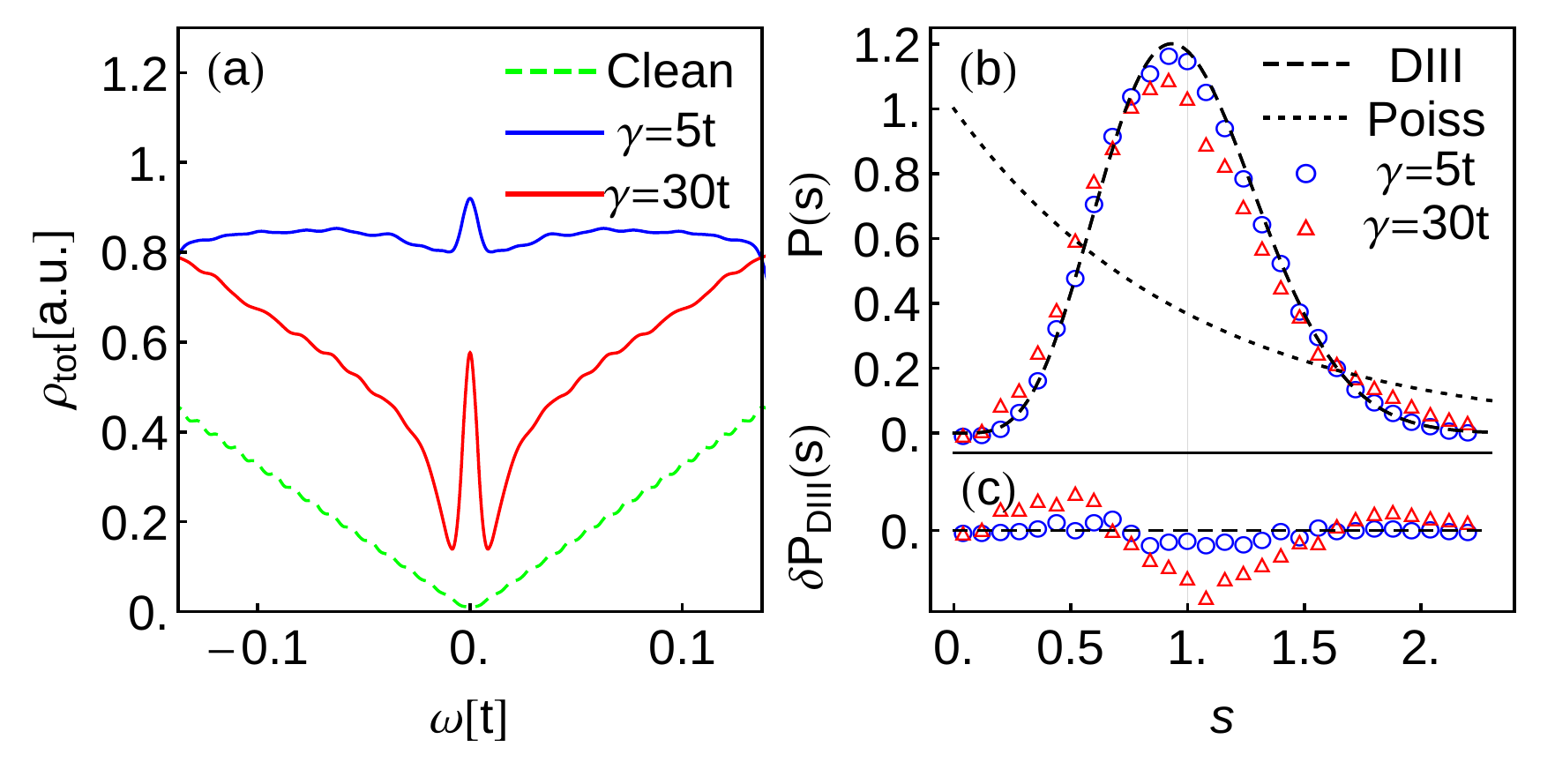}
\caption{\label{levelStat3D}   \label{DOS3Dplot}
(Color online) (a) 
Disorder-averaged total density of states $\rho_{\textrm{tot}}(\omega) $ for a three-dimensional topological NCS with disorder strength
$\gamma = 5t$ (blue solid line) and $\gamma = 30t$ (red solid line). The average is taken over $1000$ disorder configurations. 
{For comparison, the green dashed line displays the density of states of a clean Majorana cone.}
(b)~Level spacing distribution function $P(S)$ for ingap states with energies within the interval $|�\omega | < \Delta / 3$
with two different disorder strengths $\gamma = 5 t$ (blue circles) and $\gamma = 30 t$ (red triangles).
The black dashed line is the generalized Wigner surmise for class DIII (i.e., $\alpha =1$ and $\beta = 4$). The black dotted line 
represents the Poisson distribution. (c) Difference between the numerical data and the class DIII level statistics.  
 } 
\end{figure}
%%%%%%%%%%%%%%%%%%%%%%%%%%%

  % %%%%%%%%%%%%%%%%%%%%%%%%%%%%
\begin{figure*}[t]
\centering
\includegraphics[clip,angle=0,width=2.05\columnwidth]{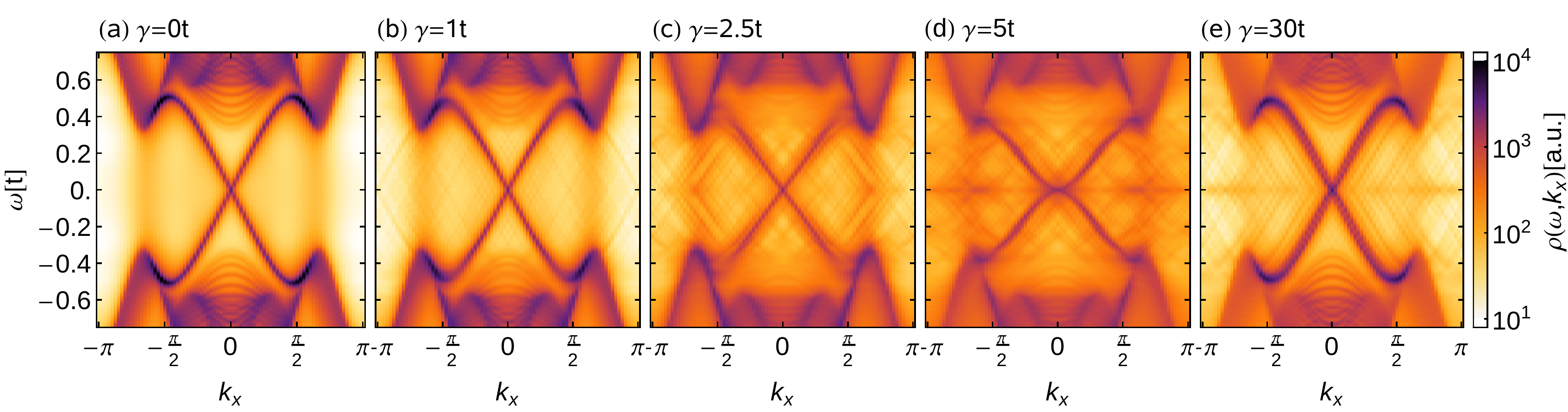}
\caption{\label{band2d}  
(Color online) 
Spectral function $A ( \omega, k_x )$ on a log scale as a function of edge momentum $k_x$ for the first three outermost layers
at the (01) edge of a two-dimensional topological NCS on a square lattice of size $80 \times 40$.  The strength of the edge disorder increases from $\gamma = 0t$ in (a)
to $\gamma = 30 t$ in (e). 
}  
\end{figure*}
%%%%%%%%%%%%%%%%%%%%%%%%%%%%

\begin{figure}[t!]
\vspace{-0.2cm}
\includegraphics[clip,angle=0,width=\columnwidth]{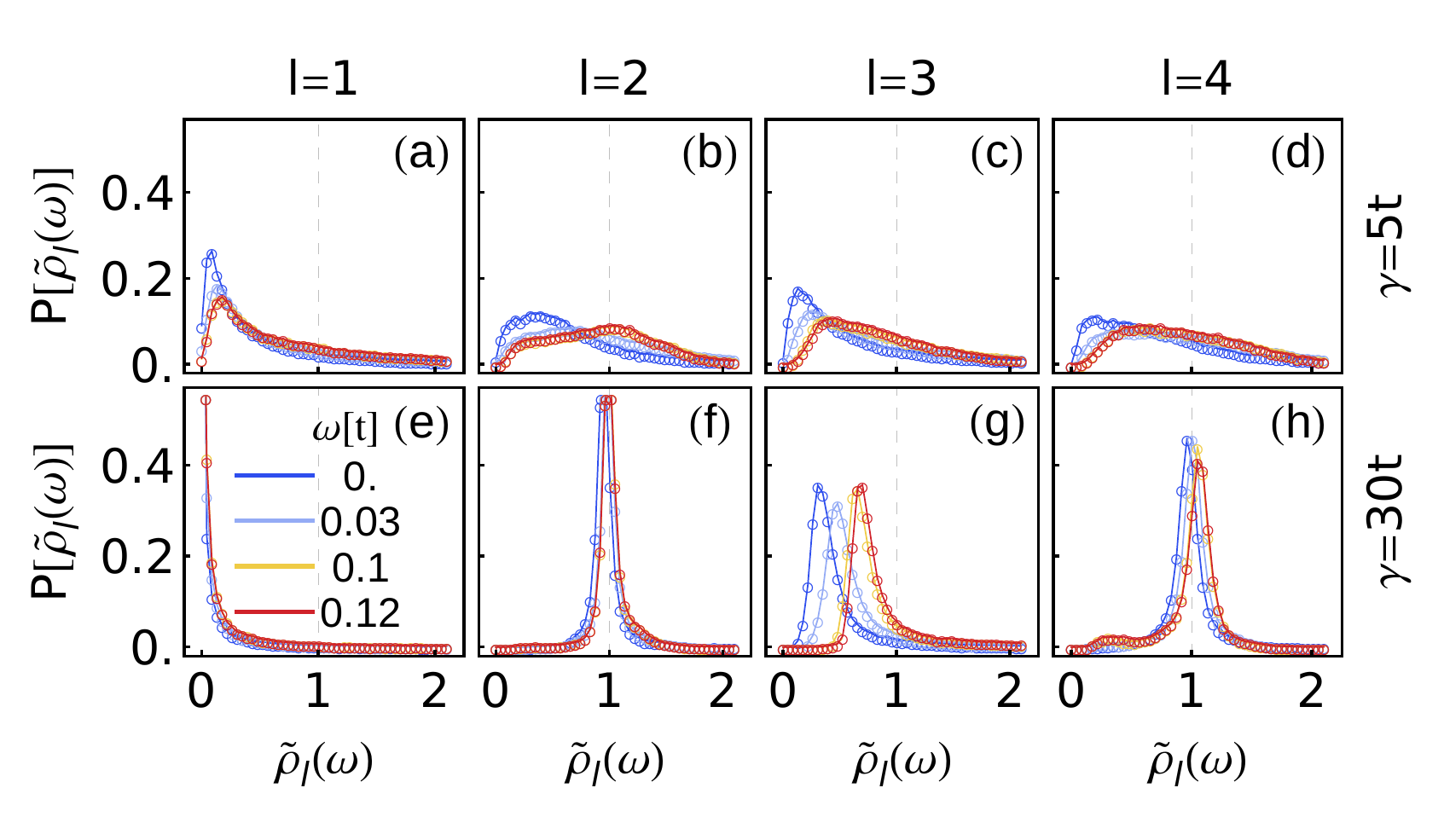}
\caption{\label{dosdist2D}  
(Color online) (a)-(h) { Layer and energy-resolved probability distribution $P[ \tilde{\rho}_l (\omega)]$ of the normalized local density of states $\tilde{\rho}_l (\omega)$
for an
ensemble of one hundred disordered two-dimensional topological NCSs with   disorder strengths $\gamma = 5 t$ and   $\gamma = 30 t$. 
The energy dependence is indicated by the color scale, with blue representing $\omega =0$ and red $\omega=\Delta/3$.
Each column shows the probability distribution $P[ \tilde{\rho}_l (\omega)]$ for a different layer.}
}
\end{figure}

Using exact diagonalization algorithms\cite{ARPACK}  we compute the
 eigenenergies $E_m$ and eigenstates $\bm{\varphi}_m$ of 
$H_{\bf k}$, Eq.~\eqref{modelDEF}, in 
the presence of surface disorder described by Eq.~\eqref{surfaceDis}.
The effects of surface impurities are best revealed in the 
local surface density of states
\begin{subequations} \label{LDOSexp}
\begin{eqnarray} \label{LDOSfun}
\rho_l  ( \omega , {\bf r}_{n}  )
=
- \frac{\hbar}{4 \pi�} \mathrm{Im} 
\sum_{j=1}^4 \sum_{m } 
\frac{ \left| \varphi_m ( l , {\bf  r}_n , j) \right|^2 }{\omega-E_m + i\eta} 
\end{eqnarray}
and the 
momentum-resolved  spectral function\cite{Schubert12}
\begin{eqnarray} \label{spectFun}
A_l  ( \omega , {\bf k}_{\parallel}  )
&=&
- \frac{\hbar}{4 \pi�} \mathrm{Im} 
\sum_{j=1}^4 \sum_{m}
\frac{ \left| \varphi_m ( l , {\bf k}_{\parallel} , j) \right|^2 }{\omega-E_m + i\eta} ,
\end{eqnarray}
\end{subequations}
with $\varphi_m ( l, {\bf k}_{\parallel} , j ) = (1 /  \sqrt{ \mathcal{N} } ) \sum_n \varphi_m ( l, {\bf r}_n , j) e^{- i {\bf k}_{\parallel} \cdot {\bf r}_n } $.
Here, ${\bf k}_{\parallel}$ denotes the surface momentum, ${\bf r}_n$ are the surface sites, $l$ represents the layer index, 
and $j$ is the combined spin and particle-hole index.
 The expressions~\eqref{LDOSexp} are evaluated  in real space with an intrinsic broadening $\eta = 0.05$ for finite lattices of size
 $50 \times 20 \times 30$ in three dimensions and  $1000 \times 40$ in two dimensions, unless otherwise stated. To compute the density of states of two-dimensional topological NCSs (Figs.~\ref{dosdist2D} and~\ref{figDOS2d}), we employ the recursive Green's function technique,\cite{Lee81}
which is numerically more efficient than direct diagonalization.

{ To obtain insight into the localization or delocalization properties of the surface states, it is useful to compute
 the probability distribution of the local density of states, $P[ \tilde{\rho}_l (\omega) ] $, and the  level spacing distribution function $P(s)$.
 The probability distribution $P[ \tilde{\rho}_l (\omega) ] $ is defined in terms of the 
 local density of states on the $l$-th layer normalized to its mean value  $\left\langle\rho_l (\omega )\right\rangle$, i.e.,\cite{Schubert10}
 \begin{eqnarray} \label{normalizedDOS}
\tilde{\rho}_l ( \omega) 
=
\rho_l (\omega )/ \left\langle\rho_l (\omega )\right\rangle  .
\end{eqnarray}
A distribution centered at $\tilde{\rho}_l ( \omega ) =1$ corresponds to extended states, while 
a distribution peaked at zero indicates localized states. 
The level spacing distribution function $P(s)$, on the other hand, is
given in terms of the normalized spacing 
 $s= \left| E_m - E_{m+1} \right| / \delta ( E_m)$ between
 two nearest levels $E_m$ and $E_{m+1}$ with $ \delta ( E_m)$ the mean level spacing near $E_m$.}
We note that for systems with a density of states that changes rapidly with energy [see, e.g., Fig.~\ref{DOS3Dplot}(a)], the normalization of the level spacing
 intervals by the mean level spacing $\delta(E_m)$ is particularly important.\cite{HaakeBook,Chou14}
 In a disordered quantum system the energy level distribution reflects the localization properties of the system:\cite{HaakeBook,Altland97} In a delocalized
 phase nearby energy levels repel each other leading to a Wigner-Dyson-like level statistics. In a localized phase, however,
  different levels can be arbitrarily close to each other, which gives rise to Poissonian  statistics.

\subsection{Three-dimensional topological NCS}\label{sec:drei3D}

We start by discussing the effects of nonmagnetic disorder on the helical Majorana states of three-dimensional topological NCSs. Figures~\ref{band3d}(a)-(e)
show the spectral function $A ( \omega, {\bf k}_{\parallel} )$ integrated over the three outermost layers,
which is of the order of the decay length of the ingap surface states. 
In the clean case, $\gamma = 0$, surface states exist at energies smaller than the bulk energy gap  $\Delta  = \Delta_{\mathrm{t} } - \Delta_{\mathrm{s}} = 1.5$
and form a helical Majorana cone, which is centered at the $\Gamma$ point of the surface Brillouin zone [Fig.~\ref{band3d}(a)].
In accordance with the discussion of Sec.~\ref{lowEnergyHams}, we find that weak and even moderately strong disorder leaves the
spectral function almost unchanged, apart from small broadening effects [Figs.~\ref{band3d}(b) and~\ref{band3d}(c)]. 
Conversely, strong surface disorder with strength of the order of the band width, $\gamma \simeq 5 t$,
completely destroys the momentum-space structure of the ingap surface states { and leads} to a large increase of the density of states in the surface layers [Fig.~\ref{band3d}(d)].
 For this disorder strength, the wave function probability densities $\left| \bm{\varphi}_m ( l, {\bm r} ) \right|^2$ exhibit sharp peaks spread through the entire layer,
see Figs.~\ref{dosdist3D}(i) and~\ref{dosdist3D}(j). 
The real-space structures of $\left| \bm{\varphi}_m ( l, {\bm r} ) \right|^2$ for different wave functions
with nearby energies are correlated, forming clusters of differently colored peaks. 
In other words, the surface state wave functions show signs of critical delocalization.
Finally, for extremely strong impurity scatterers $\gamma \gg t$, 
almost fully localized impurity states are formed in the surface layer, while in the second and third outermost layers extended states reemerge, forming helical Majorana bands
 [Figs.~\ref{band3d}(e)].\cite{Schubert12,Queiroz14}

In Figures~\ref{dosdist3D}(a)-(h) we show the layer and energy-resolved probability distribution of $\tilde{\rho}_l ( \omega)$ 
 for disorder strengths $\gamma=5t$ and $\gamma=30t$.\cite{footnote3} 
{ For $\gamma=30t$,
the probability distribution $P[ \tilde{\rho}_l (\omega) ] $ in the surface layer $l=1$ is peaked at $\tilde{\rho}_l =  0$, which signals localization,
whereas $P[ \tilde{\rho}_l (\omega) ] $ for the inward layers $l=2$ and $l=4$ has a maximum close to $\tilde{\rho}_l =  1$, indicating delocalized states.
[We note that the surface state wave function has a node in layer $l=3$ and therefore $P[ \tilde{\rho}_l (\omega) ] $ for $l=3$ is peaked closer
to  $\tilde{\rho}_l =  0$, see Fig.~\ref{dosdist3D}(g).]
With decreasing energy, the maximum of the distributions in Figs.~\ref{dosdist3D}(f) and~\ref{dosdist3D}(h) approaches $\tilde{\rho}_l =  1$, which
shows that the states with energies near zero (in particular the Majorana zero-energy states) are more extended than those
states with energies of the order of  $\Delta / 3$.  
For $\gamma \simeq 5 t$,  corresponding to the crossover from   weak to   strong disorder,
$P[ \tilde{\rho}_l (\omega) ] $ in all four layers shows a broad peak, which we interpret as a sign of critical delocalization, cf.~Figs.~\ref{dosdist3D}(i) and~\ref{dosdist3D}(j).\cite{Chou14}  }

The disorder-averaged total density of states 
$\rho_{\textrm{tot}} ( \omega ) = \sum_{l } \sum_{{\bf r}_n} \rho_l ( \omega, {\bf r}_n )$
 in Fig.~\ref{DOS3Dplot}(a)
 reveals that  for $\gamma = 5 t$ there exists a large number of ingap states, which completely fill up the superconducting gap. 
 For even larger disorder strength, $\gamma \gg t$, on the other hand, the number of ingap states 
 is reduced, indicating that the effects of disorder on the superconductor and its surface states effectively become weaker. 
 Importantly, we find that for all disorder strengths and disorder configurations there exist two extended zero-energy Majorana surface states.  These zero-energy modes appear in the total density of states of Fig.~\ref{DOS3Dplot}(a) as a narrow peak at $\omega =0$.

%%%%%%%%%%%%%%%%%%%%%%%%%%%%
%%%%%%%%%%%%%%%%%%%%%%%%%%%

Finally, we present in Fig.~\ref{DOS3Dplot}(b) the level spacing distribution function $P(s)$ for ingap states with energies within the interval 
$\left| \omega \right| < \Delta / 3$ in the presence of surface disorder with strength $\gamma = 5t$ and $\gamma = 30t$. 
Interestingly, we find that for $\gamma = 5 t$ the level statistics $P(s)$  fits the
generalized Wigner surmise for the symmetry class DIII (i.e., $\alpha =1$ and $\beta =4$).\cite{HaakeBook,Altland97,footnoteStatistics}
This indicates that the ingap states remain delocalized with significant overlap, giving rise to level repulsion.
For  $\gamma = 30 t$, however, there are deviations from the generalized Wigner surmise, which we attribute 
to the emergence of localized states in the strongly disordered surface layer $l=1$ { with energies of the order of $\Delta / 3$.}
These localized states do not exhibit level repulsion and hence lead to Poissonian level spacing statistics, which is
superimposed on the class DIII level statistics of the extended states { with energies close to zero.}
Importantly, we note that for $\gamma \gtrsim 5t$ the ingap surface states are strongly coupled to the bulk states and therefore can no longer
be described by an effective \emph{single-valley}, i.e. two-component, Dirac Hamiltonian of the form of Eq.~\eqref{3DsurfaceHam}, for which all impurity terms are forbidden by symmetry and therefore no random ensemble with a universal level spacing statistics can be achieved.

\subsection{Two-dimensional topological NCS}\label{sec:2D}

Next, we examine the helical Majorana states at the edge of a two-dimensional topological NCS
with nonmagnetic edge disorder. In Fig.~\ref{band2d} we present the spectral function $A ( \omega, k_x )$, Eq.~\eqref{spectFun},
summed over the three outermost layers. As for the three-dimensional NCS, we find that in the clean case, $\gamma=0$, 
 edge states appear at energies smaller than the bulk gap~$\Delta$,
 forming two Majorana bands that cross at  $k_x = 0$ of the surface Brillouin zone [Fig.~\ref{band2d}(a)].
As a consequence of time-reversal symmetry,  ingap states with opposite edge momenta have opposite
spin polarization.\cite{Hofmann13,brydonNJP2013,SchnyderPRL13,brydonArXiv2014} This completely prohibits backscattering
among the edge states by nonmagnetic impurities. Moreover, as discussed in Sec.~\ref{lowEnergyHams},
the only symmetry allowed perturbations of the edge Hamiltonian~\eqref{2DsurfaceHam}, are local variations of the superconducting gap $\Delta$, i.e., changes
in the Fermi velocity of the Majorana bands. Hence, one expects that nonmagnetic disorder with strength $\gamma \leq \Delta$
does not affect the surface states, which is confirmed by our numerical results in Fig.~\ref{band2d}(b).

%%%%%%%%%%%%%%%%%%%%%%%%%%%%
\begin{SCfigure}
\centering
\includegraphics[clip,angle=0,width=0.49\columnwidth]{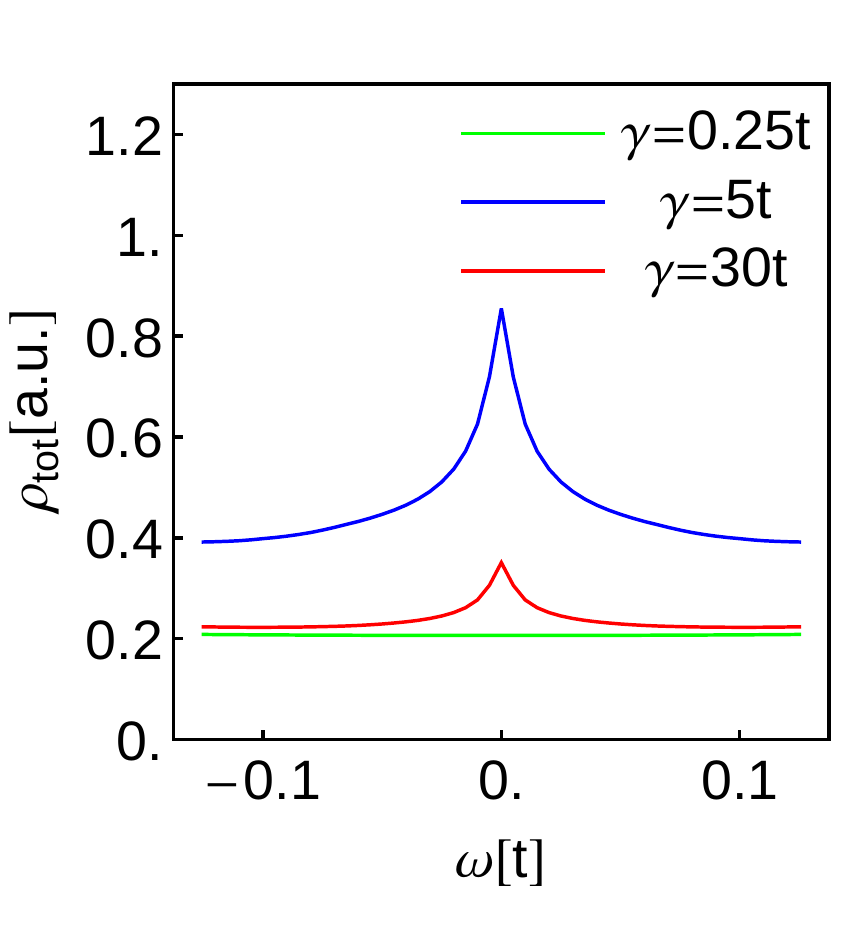}
\caption{ \label{figDOS2d}  
(Color online) 
Disorder-averaged total density of states $\rho_{\textrm{tot}} ( \omega )$ for a two-dimensional topological NCS
with disorder strengths $\gamma = 0.25t$ (green), $\gamma = 5t$ (blue), and $\gamma = 30t$ (red).
The average is taken over $100$ disorder configurations.
{\clw \scriptsize [** white space ** white space **]} }
\end{SCfigure}
%%%%%%%%%%%%%%%%%%%%%%%%%%%

For stronger edge disorder with $\gamma$ of the order of the bandwidth $5t$, the spectral function   $A ( \omega, k_x )$ becomes smeared out,
but a  momentum-space dispersion is still visible. We find that for this disorder strength, extended edge states, which form Majorana bands, strongly interact with more localized ingap states.
Moreover, comparing Figs.~\ref{band2d}(a) and~\ref{band2d}(b), we observe that 
edge disorder significantly modifies the Fermi velocity of the Majorana modes, particularly around zero energy.
Interestingly, at $\gamma = 5t$ the probability distribution $P [ \tilde{\rho}_l ( \omega ) ]$ is strongly broadened [Fig.~\ref{dosdist2D}(a)-(d)] and the
disorder-averaged total density of states $\rho_{\textrm{tot}} ( \omega ) $ exhibits a pronounced peak at $\omega =0$~(blue trace in Fig.~\ref{figDOS2d}).
The latter is reminiscent of the Dyson singularity at zero-energy which occurs in (quasi)-one dimensional dirty superconductors belonging
to symmetry class DIII.\cite{BrouwerPRL98,BrouwerPRL00,brouwerReview05}
 That is, the spectral function of Fig.~\ref{band2d}(d) and the total density of states of Fig.~\ref{figDOS2d} indicate 
that extended ingap states coexist with critically delocalized states at the edge of the superconductor.
Further increasing the disorder strength to $\gamma = 30t$, we observe that the number of ingap states decreases,
the height of the zero-energy peak in $\rho_{\textrm{tot}}$ is reduced significantly,
and the Majorana bands recover a perfectly linear dispersion (Fig.~\ref{band2d}(e) and red trace in Fig.~\ref{figDOS2d}). This shows that the effects of the
edge disorder on the bulk superconductor and its Majorana ingap states effectively decreases for $\gamma \gg t$. 
In fact, just as for the three-dimensional topological superconductor, very strong impurity scatterers
give rise to almost fully localized edges states in the surface layer,
while the extended states that form the Majorana bands are now mostly located in the second and fourth outermost layers [Fig.~\ref{dosdist2D}(e)-(h)].
Overall, we find that disorder affects the Majorana modes of two-dimensional topological NCSs less strongly than
those of three-dimensional topological NCSs. In part, this is due to the helical spin texture of the Majorana modes, 
which completely prohibits backscattering at the one-dimensional NCS edge, whereas it only partially suppresses scattering across the 
Majorana  cone of a three-dimensional topological NCS.

\section{Magnetic surface disorder}\label{sec:mag}

%%%%%%%%%%%%%%%%%%%%%%%%%%%%
\begin{figure}[t!]
\flushleft
\includegraphics[clip,angle=0,width=1\columnwidth]{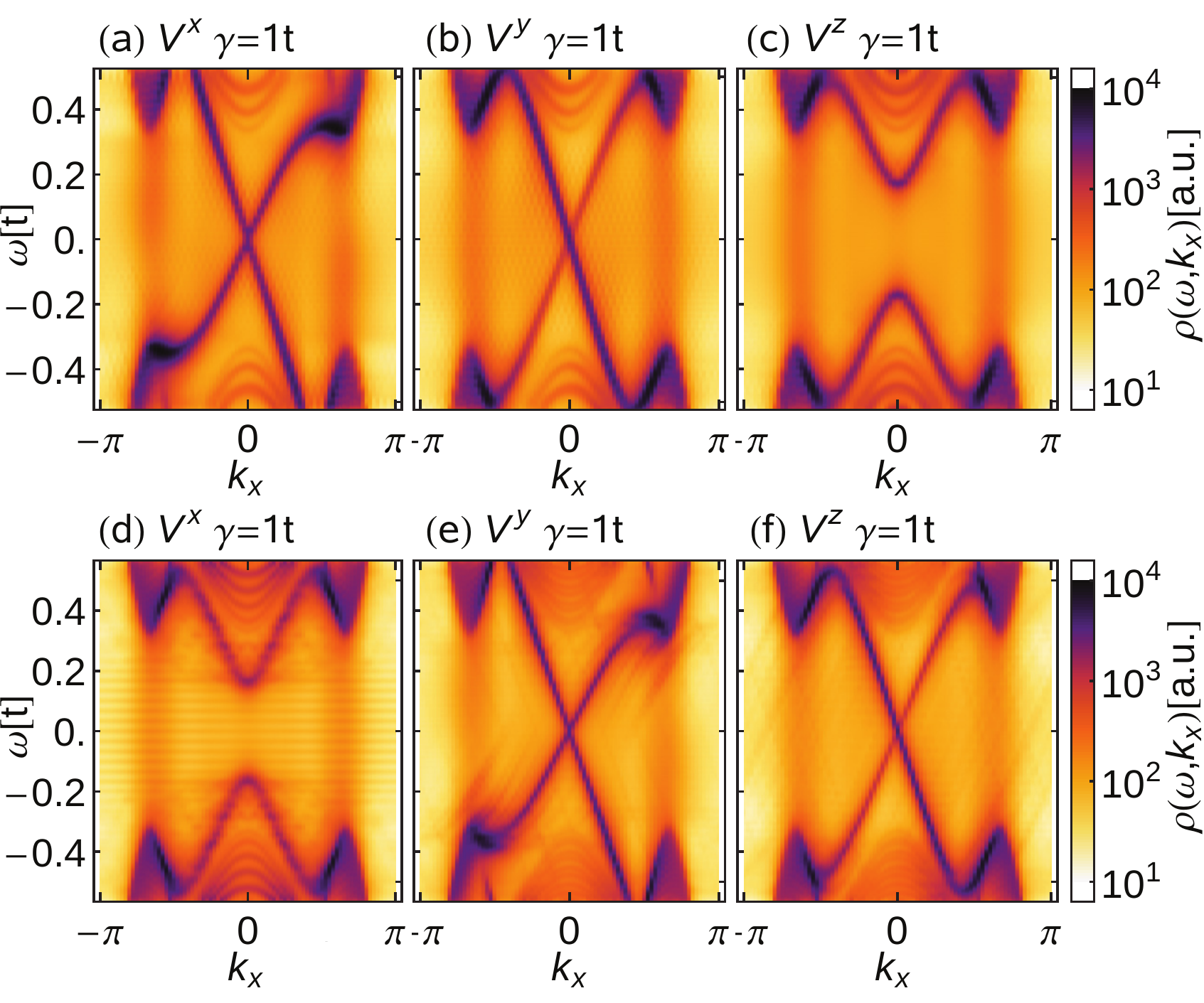}
\caption{\label{band_mag}  
(Color online) 
(a)-(c) Spectral function $A ( \omega, k_x )$ on a log scale as a function of surface momentum $k_x$ with $k_y =0$ for the first three outermost layers
at the (001) face of a three-dimensional topological NCS in the presence of magnetic surface disorder with  $\gamma = 1 t$.
In  (a), (b), and (c) the impurity spins are polarized along the $x$, $y$, and $z$ axes, respectively. 
(d)-(f) show the same as (a)-(c) but for the (01) edge of a two-dimensional topological NCS on a square lattice of size $80 \times 40$.
}  
\end{figure}
%%%%%%%%%%%%%%%%%%%%%%%%%%%

Magnetic surface disorder lifts the time-reversal symmetry protection of the helical Majorana cone  and therefore
can induce a full gap in the surface spectrum. In \mbox{Figs.~\ref{band_mag}(a)-(b)} we show the spectral function $A ( \omega, k_x)$ 
integrated over the three outermost layers at the (001) surface of a three-dimensional topological NCS in the presence
of magnetic impurities $V_{\bf r}^{x, y, z} = v ( {\bf r} ) S^{x,y,z}$, Eq.~\eqref{surfaceDis}, polarized along the $x$, $y$, and $z$ axes.
We find that whether the Majorana state is gapped out by magnetic disorder with $\gamma \leq \Delta$, depends on the direction in which the impurity spins are polarized.
As discussed in Sec.~\ref{sec:SurfDisorder},\cite{footnoteSpinPol} the Majorana bands couple most strongly to the $z$ component of the impurity spins,
since they exhibit a strong $z$-spin polarization.\cite{Hofmann13,brydonNJP2013,SchnyderPRL13,brydonArXiv2014}
Hence, $z$-polarized impurity spins open up a full gap in the surface spectrum [Fig.~\ref{surfaceDis}(c)], inducing a thermal quantum Hall state at the surface.\cite{RyuMooreLudwig2012}
On the other hand, $y$-polarized impurities only couple weakly to the Majorana cone, whereas $x$-polarized impurity spins do not affect the surface states at all, since the polarization of the Majorana bands has zero component along the $x$ axis [Figs.~\ref{surfaceDis}(a) and~\ref{surfaceDis}(b)].\cite{footnoteSpinPol2}

A similar behavior is also observed for the Majorana edge states of a two-dimensional topological NCS [Figs.~\ref{band_mag}(d)-(f)].
Here, we find that the Majorana bands couple most strongly to the $x$-spin component of the magnetic disorder [Fig.~\ref{band_mag}(d)],
whereas the $y$ component of the impurity spins does not interact with the edge states, 
which are fully polarized within the $xz$ spin plane [Fig.~\ref{band_mag}(e)].\cite{Hofmann13,brydonNJP2013,SchnyderPRL13,brydonArXiv2014}

%%%%%%%%%%%%%%%%%%%%%%%%%%%%%%%%%%%%%%%%%%%%%%%%%
%%%%%%%%%%%%%%%%%%%%%%%%%%%%%%%%%%%%%%%%%%%%%%%%%
%%%%%%%%%%%%%%%%%%%%%%%%%%%%%%%%%%%%%%%%%%%%%%%%%
\section{Conclusions and Outlook}
\label{sec:summary}
\label{sec:Conclu}

In this paper we have used large-scale exact diagonalization and the recursive Green's function technique to study the effects of strong surface disorder on the Majorana
surface bands of two- and three-dimensional topological superconductors of the symmetry class DIII.
In order to determine the effects of strong disorder, we have computed the level spacing statistics and the local density of states  of single particle wave functions.
Weak disorder with strength $\gamma$ smaller than the bulk superconducting gap $\Delta$, 
does not perturb the surface Majorana cone,
since there exists no relevant or marginal symmetry-allowed term that can couple to the surface Dirac Hamiltonian.
Very strong disorder with $\gamma$ much larger than the bandwidth, however, partially localizes the 
outermost layer while the linearly dispersive Majorana band  reappears in the second and third inward layers.
Disorder affects  states with energies close to the gap more strongly than states with energies close to zero.
In particular, our numerical data shows that for all disorder strengths and configurations two extended zero-energy states exist.
These findings suggest that no diffusive state can be realized at the edge or surface of a topological superconductor.\cite{Foster14,senthilFisherPRB00,Schubert12}
At the crossover from weak to strong disorder (i.e., for $\gamma \simeq 5 t$) 
the surface state wave functions exhibit signs of critical delocalization, particularly around zero energy.
We find that for $\gamma \simeq 5t$, the density of states of the two-dimensional topological superconductor 
diverges at zero energy, which is similar to the Dyson singularity of disordered (quasi)-one-dimensional superconductors. 

The (de)localization properties of the wave functions at the crossover from weak to strong disorder, for example the (multifractal) scaling properties, deserve further investigation. Moreover, it would be interesting to study the (de)localization properties of
weak topological superconductors or of three-dimensional topological superconductors with more than one Majorana cone (i.e., winding number $\nu > 1$). 
Our findings are of relevance for fully gapped superconductors with time-reversal symmetry and (dominant) spin-triplet pairing.
For example, Li$_2$Pt$_3$B,\cite{Yuan06,Nishiyama07} CePt$_3$Si,\cite{bauerPRL04} Cu$_x$Bi$_2$Se$_3$,\cite{sasakiAndoPRL11} Cu$_x$(PbSe)$_5$(Bi$_2$Se$_3$)$_6$,\cite{sasakiAndoArxiv14} and (Ag$_x$Pb$_{1-x}$Se)$_5$(Bi$_2$Se$_3$)$_{3y}$,\cite{FangKanatzidisPRB14} have been proposed as possible hosts of this unconventional superconducting phase.
Tunneling  experiments on disordered surfaces of these systems can be used to confirm our predictions.

\acknowledgments
The authors thank P.~Brydon, J.~Hofmann, 
P.~Ostrovsky,  C.~Timm, and P.~Wahl for useful discussions. 
%%%%%%%%%%%%%%%%%%%%%%%%%%%%%%%%%
%%%%%%%%%%%%%%%%%%%%%%%%%%%%%%%%%
%%%%%%%%%%%%%%%%%%%%%%%%%%%%%%%%%
%%%%%%%%%%%%%%%%%%%%%%%%%%%%%%%%%

\bibliography{StronDisReferences_v1}

\begin{thebibliography}{72}
\bibitem[{foo({\natexlab{a}})}]{footnote3}
\bibinfo{note}{As the ingap states are unperturbed, we cannot create a random
  ensemble to extract statistical properties. The level spacing distribution
  shows a sharp peak centered at 1, evidence of the unperturbed linear
  dispersion, as well as the density of states distribution, evidence of the
  fully extended states. This quantities are, therefore, only calculated for
  the strong disorder limit.}
\bibitem[{foo({\natexlab{b}})}]{footnoteStatistics}
\bibinfo{note}{In symmetry class DIII there is a correction to the Wigner
  surmise with $\beta=4$, due to repulsion between the levels $E_m$ and their
  images $-E_m$, see Refs.~\onlinecite{HaakeBook,Altland97}}.
\bibitem[{foo({\natexlab{c}})}]{footnoteSpinPol}
\bibinfo{note}{Note that Eqs.~\eqref{eqV3D} and~\eqref{eqV2D} are only valid in
  the limit of vanishing spin-orbit coupling, $\lambda =0$, and zero
  spin-singlet pairing amplitude, $\Delta_{\textrm{s}} = 0$.}
\bibitem[{foo({\natexlab{d}})}]{footnoteSpinPol2}
\bibinfo{note}{We expect that this finding remains valid upon inclusion of
  subdominant higher order spin-orbit coupling terms, since the out-of-plane
  component of the spin polarization of the surface states is in general
  substantially smaller than the in-plane component.}
\end{thebibliography}

 \end{document}